\documentstyle[pre,twocolumn,aps]{revtex}
\input{epsf}
\begin{document}
\draft
\title{
Ionization equilibrium of hot hydrogen plasma 
}
\author{Alexander Y. Potekhin\cite{byline}}
\address{Department of Theoretical Astrophysics, 
           Ioffe Physical-Technical Institute,
           194021, St.-Petersburg, 
           Russia
}
\date{Received 19 April 1996}
\maketitle
\begin{abstract}
The hydrogen plasma is studied at temperatures $T\sim 10^4 - 10^6$~K 
using the free energy minimization method. A simple analytic free 
energy model is proposed which is accurate at densities 
$\rho\lesssim 1$ g cm$^{-3}$ and yields convergent internal 
partition function of atoms. The occupation probability formalism is 
modified for solving the ionization equilibrium problem. The 
ionization degree and equation of state are calculated and compared 
with the results of other models. 
\end{abstract}
\pacs{PACS numbers: 95.30.Q, 95.30.T, 52.25.K}
\section{Introduction}                             % SECTION 1
\label{sect:intro}
Thermodynamic properties of astrophysical plasmas have 
been studied extensively in recent years 
\cite{ChabrierSchatzman94}. 
The theoretical models are 
based either on the physical picture or on the chemical picture 
of the plasma \cite{Ebeling-77}. 
In this paper we consider the latter. 
Our study is motivated by considerable disagreement 
among the existing models in the domain of 
partial ionization. 

We consider the simplest case of pure hydrogen plasma which shows in 
relief all specific features of the problem. This particular case is 
also important for calculating atmospheric opacities 
of degenerate stars. It is generally assumed that, due to the 
gravitational stratification \cite{AlcockIllarionov80}, outer 
spectra-forming layers of atmospheres of these stars consist of 
light elements. Observations \cite{Barstow-94} confirm that 
DA white dwarfs have virtually pure hydrogen atmospheres at 
temperatures up to 40\,000~K. 

It is worthwhile to mention two widely used 
theories for the hydrogen plasma. 
The first one has been elaborated by Hummer and Mihalas \cite{HM} 
(HM), and Mihalas, D\"appen, and Hummer \cite{MDH,Dappen-88} (MDH). 
The authors presented a detailed 
discussion of previous work and 
formulated an equation of state (EOS), using an 
occupation probability formalism to obtain a finite internal 
partition function (IPF) of atoms. 
The occupation probabilities were derived 
from the plasma microfield distribution. 
The second theory has been proposed by 
Saumon and Chabrier \cite{C90,SC91,SC92} (SC) 
(see also Ref.\ \cite{SCVH}). 
These authors have 
developed a free energy model based on effective pair potentials in 
the system of hydrogen atoms, molecules, protons, and electrons. 
The theory describes successfully either a completely ionized 
plasma \cite{C90} or neutral gas \cite{SC91} but suffers from 
difficulties in treating the partial ionization. The free energy 
models originally adopted by HM \cite{HM} and SC \cite{SC91} did not 
ensure 
reasonable convergence of IPF at high temperature and 
pressure ionization at high density. This compelled the 
authors to introduce {\em ad hoc\/} modifications \cite{MDH,SC92} 
which affected the ionization equilibrium. 
The ionization curves obtained in Refs.\ \cite{MDH} 
and \cite{SC92} at $T>10^4$~K are strikingly different. 

In this paper we study hydrogen plasma at $T\sim 10^4 - 10^6$~K when 
a considerable fraction of atoms can exist in excited states, and 
the IPF convergence becomes crucial. In Sect.~\ref{sect:chempict} we 
outline the free energy minimization method. In 
Sect.~\ref{sect:fmodel} we develop an analytic free energy model 
for weakly coupled and weakly degenerate plasma. In 
Sect.~\ref{sect:occprob} we revise the occupation probability 
formalism, and in 
Sect.~\ref{sect:results} we calculate the EOS and ionization 
equilibrium, making a comparison with the results of other models. 
\section{Chemical picture and free energy          % SECTION 2
minimization method}
\label{sect:chempict}
In the chemical picture of plasmas, 
bound objects (atoms, molecules, ions) are 
treated as elementary members of the thermodynamic ensemble, along 
with free electrons and nuclei. In the physical picture, nuclei and 
electrons (free and bound) are the only constituents of the 
ensemble. 

Both pictures can be thermodynamically self-con\-sistent, but the 
chemical picture has limited microscopic consistency. For instance, it 
does not provide a proper treatment of such cluster configurations 
as ``an atom + a close alien ion'' (a pseudomolecular ion). With 
increasing density, the ionization of an electron bound to a 
particular nucleus proceeds through a progressive delocalization 
involving cluster (``hopping'') states 
\cite{PerrotDharma-wardana94}. These  states are negligible in 
nearly ideal gases, but important at higher densities. However, 
their inclusion as new members of the thermodynamic ensemble would 
complicate the free energy model. Therefore one usually considers 
basic chemical species which dominate at low density. 

On the other hand, 
the physical picture is commonly based on diagrammatic 
expansions which converge only at low densities. 
Thus one has to resort to additional assumptions in the frames  
of either picture, in order to progress on higher density. The 
chemical picture, combined with the {\it free energy 
minimization method\/}, represents a reasonable compromise between 
the rigorous treatment and the practical application. 

The central assumption of the free energy minimization method 
\cite{Graboske-69} is the factorization of the many-body partition 
function into translational, configurational, and internal factors, 
and corresponding separation of the Helmholtz free energy $F$: 
\begin{equation}
   F(V,T,\{N_\alpha\}) = F_{\rm trans} + F_{\rm conf} + F_{\rm int}. 
\label{3.1}
\end{equation}
Here $V$ is the volume, $T$ the temperature, and $\{N_\alpha\}$ the 
set of particle numbers. The internal structure of a composite 
particle is generally affected by the surrounding, 
hence the separation (\ref{3.1}) is approximate. 

At given $V$ and $T$, the equilibrium state is determined by 
minimizing $F$ with respect to the numbers $\{N_\alpha\}$, subject 
to the stoichiometric constraints. Then the pressure $P$, 
the entropy $S$ and related quantities are obtained from $F$ 
using the well known thermodynamic relations \cite{LaLi86}. 
\section{Free energy model}                           % SECTION 3
\label{sect:fmodel}
\subsection{Plasma parameters}                       % SECTION 3.1
\label{sect:plasmapar}
Consider a plasma of electrons, protons, and hydrogen atoms. We do 
not include molecules and molecular ions, assuming that the 
temperature is high enough for their 
dissociation 
(roughly, $T > 10^4$~K; possible departure from this assumption 
will be discussed in Sect.\,\ref{sect:results}).

The charged component of the plasma is described by the coupling 
parameter $\Gamma$ and degeneracy parameter $\theta$, 
\begin{equation}
    \Gamma = \beta e^2/ a_{\rm e}, ~~~  
    \theta= T/ T_{\rm F},
\label{4.1}
\end{equation}
where $\beta=(k_{\rm B} T)^{-1}$, $k_{\rm B}$ is the Boltzmann 
constant,  
$a_{\rm e}=(\frac43\pi n_{\rm e})^{-1/3}$ 
is the mean inter-electron distance, 
$n_{\rm e}=N_{\rm e}/V$ is the electron number density, 
and $T_{\rm F}$ is the Fermi temperature  
($T_{\rm F} \approx 912\,\mbox{K}\, n_{20}^{2/3}$ and   
$\Gamma \approx [12\,500 \,\mbox{K}/ T] \, n_{20}^{1/3}$, 
where 
$n_{20} \equiv n_{\rm e}/10^{20} \,\mbox{cm}^{-3}$). 
We consider
weakly coupled and weakly degenerate plasma. The generalization to
higher degeneracy ($\theta<1$) is straightforward \cite{LaLi86}. The 
strong Coulomb coupling ($\Gamma > 1$) can be taken into account 
using the models developed in Refs.\ 
\cite{C90,Ichimaru93}.  
\subsection{Translational free energy}   % TRANSLATIONAL FREE ENERGY
\label{sect:ftrans}
The free energy of the ideal plasma $F_{\rm id} = F_{\rm trans} + 
F_{\rm int}$ is the sum of three terms produced by electrons $({\rm
e})$, protons $({\rm p})$, and H atoms. For the 
atoms and protons,
\begin{eqnarray}
&&
    \beta F^{({\rm p})}_{\rm id} / N_{\rm p} = 
    \beta F^{({\rm p})}_{\rm trans} / N_{\rm p} = 
    \ln(n_{\rm p}\lambda_{\rm p}^3) -1, 
\label{4.2}\\
&&
    \beta F^{({\rm H})}_{\rm trans} / N_{\rm H} = 
    \ln(n_{\rm H}\lambda_{\rm H}^3) -1, 
\label{4.3}
\end{eqnarray}
where $n_\alpha = N_\alpha/V$ is the number density of 
species $\alpha$, and $\lambda_\alpha = 
(2\pi\beta\hbar^2/m_\alpha)^{1/2}$ is the thermal de Broglie 
wavelength. We neglect the proton spin weight since it would yield 
an insignificant constant in the free energy. For the electrons, we 
include the spin weight and the low-density correction for the 
degeneracy \cite{LaLi86}: 
\begin{equation}
    \beta F^{({\rm e})}_{\rm id} / N_{\rm e} = 
    \ln(n_{\rm e}\lambda_{\rm e}^3/2) -1 
    + n_{\rm e} \lambda_{\rm e}^3/2^{7/2}.
\label{4.4}
\end{equation}
\subsection{Internal free energy}         % INTERNAL FREE ENERGY
\label{sect:fint}
There are different ways to define the internal free energy of atoms 
$ F^{\rm (H)}_{\rm int}$. We calculate it in the ideal gas 
approximation, neglecting interactions of atoms with surrounding 
particles. From the first principles \cite{LaLi86}, 
the ideal-gas part of the free energy is 
\begin{equation} 
    \beta F^{({\rm H})}_{\rm id} = \sum_\kappa N_\kappa  
    \left[\ln(n_\kappa\lambda_{\rm H}^3/g_\kappa) - 1 - 
    \beta\chi_\kappa \right],
\label{4.7}
\end{equation}
where $\kappa$ enumerates quantum states with statistical weights 
$g_\kappa$ and {\it non-perturbed\/} 
binding energies $\chi_\kappa$. 
It has been shown\cite{SAK95} 
that the binding energies of an atom in a plasma 
practically do not shift with increasing density 
until they merge into the continuum. 
Comparing Eqs.\ (\ref{4.7}) and (\ref{4.3}), we obtain 
\begin{equation}
    \beta  F^{\rm (H)}_{\rm int} = \sum_\kappa N_\kappa 
    \ln\left[N_\kappa/(N_{\rm H} g_\kappa 
    {\rm e}^{\beta\chi_\kappa}) \right]. 
\label{4.8}
\end{equation}
Now all nonideality effects should be included in 
the configurational term $F_{\rm conf}$.  However, $F_{\rm conf}$ 
does affect the {\em equilibrium value\/} of $ F^{\rm (H)}_{\rm 
int}$ through the distribution of the occupation numbers 
$\{N_\kappa\}$, which is {\em not\/} assumed to obey the 
ideal-gas Boltzmann law.
\subsection{Configurational free energy}    % COULOMB FREE ENERGY
\label{sect:fconf}
It is the common practice to separate 
Coulomb interactions of charged particles from interactions 
involving neutral atoms and to describe the first ones by the free 
energy of a pure Coulomb plasma.  At low density, the excess free 
energy of the charged component is given by the Debye--H\"uckel 
theory with a two-component plasma 
quantum correction \cite{DeWitt66}: 
\begin{equation}
    \beta F_{\rm C} = - {2e^3\over 3}\,\sqrt{{\pi\beta^3\over V}}\,
    (N_{\rm e}+N_{\rm p})^{3/2}
    \left( 1 - {3\,\sqrt{\pi}\over 2^{7/2}}\,\gamma \right), 
\label{4.10}
\end{equation}
where $\gamma=\frac{1}{4}(\gamma_{\rm ee}+2\gamma_{\rm ep})$, 
$\gamma_{\rm ee}$ and $\gamma_{\rm ep}$ 
being the electron-electron and electron-proton 
quantum diffraction parameters,  
$\gamma_{\rm ep} \approx \gamma_{\rm ee}/\sqrt{2} = 
2\hbar\beta e\,\sqrt{\pi (n_{\rm e}+n_{\rm p})/m_{\rm e}}$. 
Our $\gamma$ differs from the one-component plasma 
parameter $\gamma_{\rm e}$ 
employed in the SC model \cite{C90} by multiplier 0.854. 

The quantum corrections in 
Eqs.\ (\ref{4.4}) and (\ref{4.10}) allow us to extend the analytic 
free energy  model from the low density region of $\Gamma\ll 1$ and 
$\theta\gg 1$ to the moderate density, where $\Gamma\lesssim 1$, 
$\theta\gtrsim 1$. We have checked 
(see also Fig.\,12 of Ref.\ \cite{C90}) 
that the account of 
these corrections extends the validity range of the model 
by more than an order of magnitude toward higher density 
or lower temperature. 

The neutral component produces additional configurational terms, 
which describe interactions of neutral species with neutral and 
charged particles. For atoms and ions, these interactions are 
described by pair potentials, while the electrons are assumed to 
adjust to any configuration of the heavy particles. A free 
energy model based on this approach has been elaborated by SC 
\cite{SC91,SC92} and extended recently to helium plasma 
\cite{AparicioChabrier94}. Our modified version of the pair 
potential free energy model $F_{\rm pair}$ is given in Appendix. 
However, the contribution to the free energy, which comes from 
unbounding of atoms in the course of their interactions with other 
particles, is not fully griped at this approach. 
Namely, the effective hard sphere diameters of atoms, 
derived from effective binary potentials, 
appear to be too small to enforce ionization 
in high-$T$ and high-$\rho$ domain. 
Below we consider the problem in more detail. 
\subsection{Pressure destruction of atoms}  % PRESSURE IONIZATION
\label{sect:prion}
The central problem of calculating the occupation numbers 
$\{N_\alpha\}$ is to achieve the self-consistent convergence of IPF, 
that is most difficult in the case of partial ionization. 
There have been many attempts to solve the problem; 
some of them are discussed in Ref.\ \cite{HM}.

Clearly, a bound state cannot be populated, if the corresponding 
``size'' of the electronic ``orbit'' (the electron cloud) is too 
large. In other words, the electron cannot be bound if there is not 
enough free space, or if it suffers from strong perturbations. The 
problem is how to include the unbounding into the chemical picture 
in a self-consistent manner. 

The simplest way to obtain the convergence is to truncate the IPF at an 
appropriate ``critical'' quantum number $\kappa^\ast$, for example, 
corresponding to a quantum-mechanical size of an atom 
$l_{\kappa^\ast}$ comparable with the mean interparticle distance 
\cite{Herzfeld16}. However, since $\kappa^\ast$ depends on the  
physical conditions (particularly, on density), the abrupt cutoff 
produces unrealistic discontinuities in the free energy. Continuous 
truncation procedures imply introduction of {\it occupation 
probabilities\/} $w_\kappa$ into the IPF, which suppress higher 
states and ensure the convergence. However, as has been shown by 
Fermi \cite{Fermi24} and emphasized by HM \cite{HM}, the 
introduction of $w_\kappa$ requires a modification of the free energy. 
Therefore $w_\kappa$ should be consistent with the adopted form of 
$F_{\rm conf}$. 

In the free energy model $F_{\rm pair}$ (Appendix), the repulsive 
interparticle interactions are simulated by repulsion of hard 
spheres (HS) with appropriate diameters $d$. However, for a high $T$ 
or for a high atomic level, $d$ becomes much smaller than the atomic 
size $l_\kappa$, so that other plasma particles can 
penetrate the atom. If this happens, then the electronic orbit 
gets embedded into the plasma, which screens the attraction to 
the nucleus and makes the electron unbound. Then the atomic 
constituents should be extracted from the neutral component and 
treated as independent participants in the charged component. 
Similar situation (known as {\em pressure ionization\/}) occurs even 
for ground state atoms, if the density is high enough. 

To allow for the plasma screening of the intra-atomic Coulomb 
interactions, the static screened Coulomb potential (SSCP) has been 
widely used (e.g., Ref.\,\cite{FGVH}). However this approach has 
been strongly criticized \cite{HM}. 
More consistent is the dynamical screening 
recently investigated in the framework of the thermodynamic Green 
function method \cite{SAK95}. However, the 
practical need (e.g., in astrophysics) for 
large arrays of thermodynamic data 
necessitates looking for a simplified approach. 

A convenient model has been described by HM. 
The configurational term given by
Eq.~(4.72) of Ref.~\cite{HM} for the hydrogen plasma may be 
represented as 
\begin{equation}
    F^{\rm (HM)}_{\rm conf} = k_{\rm B}T \sum_\kappa N_\kappa 
    ( n_{\rm H} \tilde{v}_{1\kappa} + 
    n_{\rm p} \tilde{v}_{{\rm p}\kappa} ), 
\label{4.28b}
\end{equation}
where $\tilde{v}_{1\kappa}$ and $\tilde{v}_{{\rm p}\kappa}$ are 
characteristic volumes associated, respectively, with the atom-atom 
and atom-ion interactions. The first term includes the interaction 
of an atom in a state $\kappa$ with the ground state atom only. 
This is the ``low excitation approximation'' proposed by HM to
make $F_{\rm conf}$ linear in $N_\kappa$, which was essentially 
employed in deriving the occupation probabilities (despite the fact 
that even at this approximation the linearity obviously breaks down 
for the ground state atoms). For purely neutral gas ($n_{\rm 
p}=0$), Eq.~(\ref{4.28b}) reproduces the free energy derived by 
Fermi \cite{Fermi24}. The atom-ion interaction volume 
$\tilde{v}_{{\rm p}\kappa}$ in this approach 
is due to microfield perturbations; it 
depends on the principal quantum number 
in a complicated way, but it is always 1--2 orders of magnitude 
larger than $\tilde{v}_{1\kappa}$. For example, for the ground-state 
hydrogen, Eq.\,(4.70) of Ref.\,\cite{HM} yields 
$\tilde{v}_{{\rm p}1}=128\,(4\pi/3)a_{\rm B}^3$, where $a_{\rm B}$ 
is the Bohr radius.

It was found \cite{MDH} that this model does not provide pressure 
ionization at high density, which is not surprising. Indeed, since 
$\tilde{v}_{{\rm p}\kappa} \gg \tilde{v}_{1\kappa}$, $F_{\rm 
conf}^{\rm (HM)}$ may increase more rapidly with growing $n_{\rm p}$ 
than with $n_{\rm H}$, shifting the equilibrium towards {\em 
lower\/} ionization degree at {\em higher\/} densities. In order to 
ensure the desired ionization, MDH introduced an artificial 
``pressure ionization term'' $F_5$ into the free energy, which 
rapidly increased whenever the density of neutral fraction exceeded 
$10^{-2}$ g cm$^{-3}$. 
It can be shown that it is this term 
(and not the inaccurate HS treatment, as supposed in 
Ref.\ \cite{SCVH}) that produces 
an unrealistically stiff EOS at $\rho>10^{-2}$~g~cm$^{-3}$. 

As argued in Refs.\ \cite{Herzfeld16,Fermi24,HM}, the 
quantum-mechanical 
atomic size $l_\kappa$ [Eq.\,(\ref{4.21})] 
should not exceed the mean interparticle distance. 
This is not a problem at low temperature, when the effective HS 
diameters $d_{\kappa\kappa'}$ are larger than $l_\kappa$. 
At high $T$, however, $d_{\kappa\kappa'}$ become small, 
allowing configurations with strongly overlapping 
wavefunctions of neutral atoms. 
SC escaped this difficulty by introducing an 
additional hard core [Eq.~(14) of 
Ref.~\cite{SC92}] in the effective potentials. 
However, a large hard core inside the atom seems to be 
unrealistic. We propose a modification of 
the free energy, which has another interpretation. 

An electron can be treated as bound to a particular nucleus, if 
only its wave function does not overlap strongly with wave functions 
of other electrons, either free or bound to neighboring atoms. From 
the classical point of view, the atomic electron becomes unbound 
when another electron falls inside its orbit and shields the 
attraction to the nucleus. This resembles the plasma screening of 
the nucleus in the SSCP model but does not imply the 
collective nature of the screening. 

For a given state $\kappa$, the probability that such unbounding 
does {\it not\/} occur can be estimated at low density 
from the Poisson distribution, 
$
    p_\kappa = \exp[-n_{\rm c}\,v_\kappa], 
$
where 
$n_{\rm c} = (n_{\rm e}+n_{\rm H})$ 
is the total number of randomly distributed 
electronic clouds (including those which 
are bound to nuclei), and 
$v_\kappa = \frac43\pi l_\kappa^3$. 
The unbounding requires to exclude the overlapping configurations  
from the total partition function, thus reducing the volume of the 
phase space available to the system. Equivalently, the existence of 
an atom in the state $\kappa$ corresponds to an event with 
probability $p_\kappa$ and thus diminishes the entropy. The total 
negentropy corresponding to a set of occupation numbers 
$\{N_\kappa\}$ is $-\sum_\kappa N_\kappa \ln p_\kappa$, which gives
the free energy contribution
\begin{equation}
    F_{\rm ub} = k_{\rm B}T 
    N_{\rm c} n_{\rm H} \bar{v}_{\rm H}, 
\label{Fub}
\end{equation}
where 
$\bar{v}_{\rm H} = \sum_\kappa N_\kappa v_\kappa / N_{\rm H}$ 
is the average atomic volume. 

A similar term has been introduced by HM who, however, considered 
the destruction of atoms by microfields fluctuating 
due to the motions of surrounding heavy particles 
(as discussed in Sect.~IV{\it b}(ii) of Ref.\ \cite{HM}). 
Although we 
readily agree that nearby passages of positive ions can 
ionize a particular atom, this process does not affect the 
occupation numbers at the thermodynamic equilibrium, since it is 
compensated by the inverse (neglected) process, 
owing to the principle of detailed balance. 
The net effect of both processes is not given in advance 
but itself should be determined from the thermodynamic 
equilibrium conditions. 
In contrast, the unbounding by an excessive 
negative charge occurring inside an electron orbit has no balancing 
counterpart. Moreover, it seems inconsistent to include any 
dynamical process dependent on particle momenta into $F_{\rm conf}$ 
after separating the translational term $F_{\rm trans}$, 
because the separation (\ref{3.1}) implies that the other terms may 
depend only on particle configuration coordinates. For this 
reason we do not consider atomic collisions with ions \cite{HM} and 
free electrons \cite{MM79}.

Additional arguments in favor of the modification 
(\ref{Fub}) of the excess free energy (\ref{4.28b}) 
will be given in Sect.~\ref{sect:optprob}. 

Finally, the total free energy
\begin{equation}
    F = F_{\rm id} + F_{\rm C} + F_{\rm pair} + F_{\rm ub} 
\label{4.29}
\end{equation}
is given by Eqs.\ (\ref{4.2}), (\ref{4.4}), (\ref{4.7}), 
(\ref{4.10}), (\ref{Fub}), and (\ref{A1}). 

Other chemical species can be easily included in our 
analytic model. For example,  H$_2$ molecules can be 
taken into account by adding 
van der Waals and HS terms with appropriate effective diameters
in Eqs.\ (\ref{4.21b}), (\ref{4.25}) and (\ref{4.27}). 
The generalization of the unbounding term is 
\begin{equation}
    F_{\rm ub} = k_{\rm B}T 
    N_{\rm c} (n_{\rm H} \bar{v}_{\rm H} + 
    n_{\rm H2} \bar{v}_{\rm H2}), 
\label{FubH2}
\end{equation}
where $n_{\rm c} = n_{\rm e}+n_{\rm H}+n_{\rm H2}$, 
and $\bar{v}_{\rm H2}$ is the average molecular volume. 
\section{Occupation probability formalism} % OCCUPATION PROBABILITY
\label{sect:occprob}
\subsection{Generalized Saha equation}              % SECTION 4.1
\label{sect:saha}
The internal free energy (\ref{4.8}) can be rewritten in a more 
familiar form. Let us replace the Boltzmann 
distribution $N_\kappa\propto g_\kappa \exp(\beta\chi_\kappa)$ 
(which yields the divergent ideal IPF) by any real distribution 
\begin{equation}
N_\kappa = N_{\rm H} w_\kappa g_\kappa \exp(\beta\chi_\kappa)/Z_w, 
\label{real:distr}
\end{equation}
where the generalized IPF 
\begin{equation}
    Z_w = \sum_\kappa g_\kappa w_\kappa {\rm e}^{\beta\chi_\kappa} 
\label{4.8b}
\end{equation}
plays role of a normalization constant. 
Then 
\begin{equation}
    \beta F^{({\rm H})}_{\rm int} = 
    \sum_\kappa N_\kappa \ln w_\kappa - N_{\rm H} \ln Z_w. 
\label{4.8a}
\end{equation}
Note that $w_\kappa$ and $Z_w$ in 
Eqs.\ (\ref{real:distr})--(\ref{4.8a}) 
can be multiplied by a common factor. 
It does not affect the $N_\kappa / N_{\rm H}$ distribution, 
but should be chosen consistent with the ionization 
equilibrium conditions. 

Although the first sum in Eq.~(\ref{4.8a}) is naturally derived
from the rigorous Eq.~(\ref{4.7}), 
it was often omitted in the internal free energy \cite{SC91} 
or regarded as a part of $F_{\rm conf}$ \cite{HM}. 
Meanwhile, it has a clear physical meaning: $-N_\kappa \ln 
w_\kappa$ is a contribution to the ideal-gas part of the entropy 
due to the  
correction $w_\kappa$ to the probability that $\kappa$th 
state is occupied. 
The factors $w_\kappa$ are traditionally called occupation 
probabilities, although they do not always have direct probability 
meaning. 

The minimum of the Helmholtz free energy under the stoichiometric 
constraints requires 
\begin{equation}
    \partial F/\partial N_\kappa = \partial F/\partial N_{\rm p} + 
    \partial F/\partial N_{\rm e} . 
\label{5.1}
\end{equation}
Separation of $F_{\rm id}$ from $F_{\rm conf}$ allows one to 
rewrite Eq.~(\ref{5.1}) in the form of the Saha equation: 
\begin{equation}
    n_\kappa = n_{\rm p} n_{\rm e} (\lambda_{\rm p} \lambda_{\rm e} 
    / \lambda_{\rm H})^3 w_\kappa (g_\kappa /2)
    \exp[\beta(\chi_\kappa + \Lambda_{\rm deg})], 
\label{5.2}
\end{equation}
where $\Lambda_{\rm deg}=n_{\rm e}\lambda_{\rm e}^3/2^{5/2}$ is the 
correction due to the partial electron degeneracy, and 
$w_\kappa$ is defined by 
\begin{equation}
    k_{\rm B} T \ln w_\kappa = {\partial F_{\rm conf} \over 
    \partial N_{\rm p}} + {\partial F_{\rm conf} \over \partial 
    N_{\rm e}} - {\partial F_{\rm conf} \over \partial N_\kappa}. 
\label{5.3}
\end{equation}
This definition is consistent with 
Eqs.\ (\ref{real:distr})--(\ref{4.8a}), 
and it fixes the above mentioned common factor. 
These occupation probabilities have the same meaning as those 
considered by HM but take into account charged 
particles. Therefore, Eq.~(\ref{5.3}) {\it generalizes\/} Eq.~(2.18) 
of Ref.\ \cite{HM} to the case when ionization-recombination 
processes are allowed in the system. 

Equivalently, one may adhere the traditional definition, 
$
    \ln {w}_\alpha = 
     - \beta \, \partial F_{\rm conf} / \partial N_\alpha,
$
and replace $w_\kappa$ in Eq.~(\ref{5.2}) by 
${w}_\kappa / ({w}_{\rm e} {w}_{\rm p})$. 
However, since proton and electron cannot be destroyed 
by the external fields, we put their 
``occupation probabilities'' equal to unity, 
thus choosing the definition (\ref{5.3}) 
for the atomic occupation factors $w_\kappa$. 

An equation equivalent to Eqs.\ (\ref{5.2}) and (\ref{5.3}) was 
derived by Fontaine {\it et al.\/}\ \cite{FGVH} (FGVH), who however 
did not introduce the occupation probabilities explicitly. 

If all $w_\kappa$ were known, then Eq.~(\ref{5.2}) would give 
direct solution to the problem. However, since $F_{\rm conf}$ 
depends on the occupation numbers, Eqs.\ (\ref{5.2}) and (\ref{5.3}) 
are to be solved together. Nevertheless, this reformulation of the 
problem is useful, because the coupled equations can be solved 
iteratively. First, one chooses an initial value of 
$w_\kappa$ and calculates 
the particle numbers $\{N_\alpha\}$ from Eq.~(\ref{5.2}). Then 
$w_\kappa$ are refined by substituting $\{N_\alpha\}$ into 
Eq.~(\ref{5.3}). At low densities, where many excited states are 
populated, this procedure appears to be more efficient than 
alternative schemes \cite{MDH,Graboske-69,FGVH}. 

Substituting Eqs.\ (\ref{4.29}) and (\ref{A1}) into Eq.~(\ref{5.3}) 
we decompose $w_\kappa$ into five factors, 
\begin{equation}
    w_\kappa = w^{(\rm ub)}_\kappa w^{\rm (C)} 
    w^{\rm (HS)}_\kappa 
    w^{({\rm H})}_{\kappa} w^{\rm (in)}_{\kappa} ,
\label{5.4}
\end{equation}
corresponding to the unbounding of atoms, the Coulomb 
interactions of charged particles, the hard-sphere repulsion, and 
the corrections due to atom-atom and atom-ion 
attraction. Equation (\ref{Fub}) yields
\begin{equation}
    w^{(\rm ub)}_\kappa = \exp[ -  n_{\rm c} v_\kappa]. 
\label{5.7b}
\end{equation}
Expanding $F_{\rm C}$ and $F_{\rm pert}$ in powers of particle 
numbers, keeping quadratic terms, and using Eq.~(\ref{5.3}), 
we obtain
\begin{eqnarray}
    \ln w^{\rm (C)} &=& 
    - \sqrt{4\pi (\beta e^2)^3 (n_{\rm e} + n_{\rm p})} 
    \left( 1 - \sqrt{\pi/8}\,\gamma \right), 
\label{5.5} \\ 
    \ln w^{({\rm H})}_\kappa &=& 2\beta 
    \sum_{\kappa'} n_{\kappa'} \, a_{\kappa\kappa'} , 
\label{5.6} \\ 
    \ln w^{\rm (in)}_\kappa &=& \beta n_{\rm p} a_\kappa -   
    \beta \sum_{\kappa'} n_{\kappa'}\, a_{\kappa'} ,
\label{5.7}
\end{eqnarray}
where $a_{\kappa\kappa'}$ and $a_{\kappa}$ are the van der Waals 
constants defined in Appendix. 
Our calculations show that the factors $w^{\rm(in)}$ 
and $w^{\rm (H)}$ are close to unity, being, thus, unimportant. 
In contrast,  $w^{\rm (C)}$ is significantly less than unity at 
high densities, even at a relatively low ionization. 

For the HS repulsion, Eq.~(\ref{4.28}) gives
\begin{equation}
    \ln w^{\rm (HS)}_\kappa = {
    (\ln w^{(0)}_\kappa)\,(1-\eta/2) -5\eta^2+3\eta^3 
    \over (1-\eta)^3},
\label{5.7a}
\end{equation}
where $\eta$ is the filling factor, and 
\begin{equation}
    \ln w^{(0)}_\kappa = - {4\pi\over3} \left[ \sum_{\kappa'} 
    n_{\kappa'} \, (d_{\kappa\kappa'}^3 - d_{\kappa'}^3) 
    + n_{\rm p}\,d_\kappa^3 \right]. 
\label{5.8}
\end{equation}
If $\eta \ll 1$, then $w^{\rm (HS)}\approx w^{(0)}$. In practice, 
the non-linear corrections given by Eq.~(\ref{5.7a}) may be 
important. We always have $w^{\rm (HS)}<1$, since $d_{\kappa'} < 
d_{\kappa\kappa'}$ in our model. 

The formalism can be generalized, e.g., for 
formation of molecules. 
The dissociation--recombination 
equilibrium is given by 
\begin{equation}
    n_{\rm H2} = n_{\rm H}^2 (\lambda_{\rm H} \sqrt{2})^3 
    Z_{w2} / Z_w^2, 
\label{disseq}
\end{equation}
where 
$Z_{w2}$ 
is the internal molecular partition function, generalized 
through multiplying each $\mu$th term by an occupation probability 
$w^{\rm (H2)}_\mu$ (e.g., Eq.~(21) of Ref.\ \cite{SC91}). 
From the equation of chemical equilibrium 
\begin{equation}
    \partial F/\partial N_{\rm H2} = 
    2 \,(\partial F/\partial N_{\rm p} + 
    \partial F/\partial N_{\rm e}) ,
\label{chemeqH2}
\end{equation}
we conclude 
that the molecular occupation probability can be defined as 
\begin{equation}
    k_{\rm B} T \ln w^{\rm (H2)} = 2\left( 
    {\partial F_{\rm conf} \over 
    \partial N_{\rm p}} + {\partial F_{\rm conf} \over \partial 
    N_{\rm e}} \right)
    - {\partial F_{\rm conf} \over \partial N_{\rm H2}}. 
\label{wH2}
\end{equation}
Keeping the traditional definition 
of $w$ both for atoms and for molecules 
would result in the same Eq.~(\ref{disseq}). 

For the perturbation and HS factors in $w^{\rm (H2)}$, 
Eqs.\ (\ref{5.6})--(\ref{5.8}) remain valid  
if to use relevant scaling factors; 
the Coulomb factor $w^{\rm (C)}$ should be squared; 
and 
the unbounding factor is derived from Eq.~(\ref{FubH2}): 
\begin{equation}
    \ln w^{\rm (H2,ub)}_\mu = 
    - n_{\rm H} (v_{\mu}-\bar{v}_{\rm H})
    - n_{\rm H2} (v_{\mu} - \bar{v}_{\rm H2})
    - n_{\rm e} v_{\mu}.
\label{wubH2}
\end{equation}
As a test example (although marginal to the present discussion), 
we have implemented this approach to molecular formation, 
utilizing a simplified treatment of $Z_{w2}$ \cite{RRN95,Ediss}. 
In this case, the unbounding factor (\ref{wubH2}) 
turned out to be unimportant, since 
the degree of dissociation is mainly determined by relation 
between atomic and molecular effective HS diameters. 

The situation is very different for the ionization of atoms. 
The repulsion 
factor $w^{\rm (HS)}$ becomes important at moderate $\rho$ 
and low $T$, but 
$w^{\rm (ub)}$ strongly affects mostly excited states 
at any density (the ground state at high density), 
decreasing rapidly as the conditions for state survival become 
violated. 
This leads to a physically reasonable convergence of the IPF 
at low densities (thus radically improving the ionization degree at 
$T\gtrsim 10^{4.5}$~K) and pressure ionization at high densities. 
\subsection{Thermodynamic and optical occupation probabilities} 
%                                         THERMODYNAMICS AND OPTICS 
\label{sect:optprob}
The occupation probability technique described above allows one to
calculate thermodynamic properties of partially ionized hydrogen
plasma. However, it is still insufficient for describing optical
properties of the plasma.

The free energy model presented in Sect.~\ref{sect:fmodel} and in 
Appendix allows for close configurations of atoms with protons, 
because the atom-ion repulsion diameters $d_\kappa$ do not exceed 
the quantum-mechanical sizes of atoms. As argued in 
Sect.~\ref{sect:chempict}, the close configurations simulate 
cluster states (approximately treated as interacting atoms and ions 
in the frames of the chemical picture). This approach yields 
physically plausible EOS. On the other 
hand, specific quantum-mechanical properties (e.g., frequencies and 
oscillator strengths of radiative transitions) of clusters most 
likely differ from those of isolated atoms. Formation of many 
different close configurations should manifest itself in optics as 
{\it quasicontinuum\/}.

Therefore, one should discriminate between the {\it thermodynamic 
continuum\/} (the states which do not contribute to the generalized 
IPF), and the {\it optical continuum\/} (the states strongly 
perturbed by surrounding). This dichotomy was first realized by 
Rogers \cite{Rogers86}, who developed the concept of the optical 
and plasma continua using the physical picture. A correct account 
of the quasicontinuum has been also taken in a recent study of 
line shapes in hydrogen opacities \cite{StehleJacquemot93}. 

The optical continuum can be determined from consideration of Stark 
merging of spectral lines of an atom affected by plasma microfields 
\cite{IT}. This leads to the 
atomic ``survival'' probabilities $\tilde{w}_\kappa$, generally 
different from $w_\kappa$ introduced in thermodynamics. Let us call 
$\tilde{w}_\kappa$ the {\it optical\/} occupation probability, to 
avoid confusion with $w_\kappa$. The occupation probabilities based 
on the plasma microfield distribution 
\cite{HM,MDH,Dappen-88} are in fact the optical ones. Their 
implication in thermodynamics leads to physically unrealistic EOS 
(cf.\ Ref.\ \cite{SCVH}) 
due to the incorrect treatment of close configurations in the free 
energy.  Ionization equilibrium would be equally implausible without 
an {\it ad hoc\/} ``pressure ionization'' term \cite{MDH}. Indeed, 
substituting the excess free energy (\ref{4.28b}) into 
Eq.~(\ref{5.1}) we would arrive at Eq.~(\ref{5.2}) with $w_\kappa$ 
replaced by 
\begin{equation}
    w^{\rm (HM)}_\kappa = \exp\left[ 
    - n_{\rm H} \tilde{v}_{1\kappa} - 
    n_{\rm p} \tilde{v}_{{\rm p}\kappa} +
    \sum_{\kappa'} n_{\kappa'} 
    (\tilde{v}_{{\rm p}\kappa'} - \tilde{v}_{1\kappa'}) \right]. 
\label{5.9}
\end{equation}
Since $\tilde{v}_{{\rm p}\kappa} \gg \tilde{v}_{1\kappa}$, the last 
(positive) term in Eq.~(\ref{5.9}) may dominate and yield the 
``occupation probabilities'' 
which grow exponentially with $n_{\rm H}$. 
Then one would get pressure neutralization instead of pressure 
ionization at high densities 
\cite{Pavlov93}. 

In contrast, Eq.~(\ref{Fub}) leads to the occupation 
probabilities, Eq.~(\ref{5.7b}), which decrease exponentially with 
density and produce the desired pressure ionization.

The occupation probabilities given by Eq.~(4.71) of Ref.\ \cite{HM} 
can be presented (for pure hydrogen plasma) in the form (\ref{5.9}) 
but {\em without the last term\/}. This form of $w_\kappa$ cannot be 
rigorously derived from Eq.~(\ref{4.28b}).  Note, however, that the 
leading factor $n_{\rm p}\tilde{v}_{{\rm p}\kappa}$ in the exponent 
occurs due to the Inglis--Teller effect \cite{IT} 
which is optical but not 
thermodynamic. Accordingly, these results can be used for 
calculating the fraction of atoms which are only slightly perturbed 
by plasma microfields so that they are able to contribute to 
the atomic opacities. For this purpose, we use an expression similar 
to that in Ref.\ \cite{HM}. However we take into account that 
$F_{\rm pair}$ includes the HS term, which 
implies that the distance between ions and atoms cannot be shorter 
than $d_\kappa$. Thus the HS volume should be subtracted from the 
interaction volume $\tilde{v}$. The latter one has been estimated by 
several authors using different (not always justified) 
approximations, as discussed by HM \cite{HM}. A reasonable 
order-of-magnitude estimate reads 
$\tilde{v}=\frac43\pi(4l_\kappa)^3$. For 7 lowest states this 
estimate is intermediate between more complicated 
Eqs.\ (4.69) and (4.70) of Ref.\ \cite{HM}, and for the ground state 
atom it fits the latter exponential with an accuracy of 9\%. 
Finally, we adopt 
\begin{equation}
    \tilde{w}_\kappa = \exp\left[ - \case{4\pi}{3} n_{\rm p} 
    \left((4l_\kappa)^3 - d_\kappa^3(T)\right)\right]. 
\label{optprob}
\end{equation}
The above considerations emphasize that $\tilde{w}_\kappa$ 
determine solely the optical properties of the plasma 
and they should not be used in the construction of the IPF. 
\section{Results and discussion}                   % SECTION 5
\label{sect:results}                               % RESULTS
\subsection{Ionization equilibrium}
\label{sect:ionres}
The free energy model described in Sect.~\ref{sect:fmodel} has been 
applied to calculation of the thermodynamic properties of plasma 
using the method of Sect.~\ref{sect:occprob}. The 
ionization isotherms are shown in Fig.~1. Light solid curves 
represent the fraction of all H atoms, $f_{\rm H} = n_{\rm 
H}/(n_{\rm H}+n_{\rm p})$, and dashed lines display the fraction of 
ground state atoms. The results are in general agreement with Ref. 
\cite{SC92} but disagree with Ref.\ \cite{MDH}. If, for example, 
$\rho=0.1$ g cm$^{-3}$, Fig.~2 of Ref.\ \cite{MDH} shows practically 
zero ionization at $T\leq 10^{4.5}$~K, whereas according to Ref. 
\cite{SC92} there is a considerable amount (about 6\% by mass) of 
free protons at $T=22\,000$~K. Our result coincides with the latter 
one. The pressure ionization in our Fig.~1 proceeds smoothly at 
high densities, again in agreement with Ref.\ \cite{SC92}, 
but contrary to almost abrupt pressure ionization of Ref.\ 
\cite{MDH}.

Thick solid and dashed lines in Fig.~1 are obtained using 
Eq.~(\ref{optprob}) and show the fraction of those atoms whose 
optical properties are not destroyed by plasma microfields, and 
which therefore should be used in the opacity calculations. At $\rho 
\lesssim 10^{-3}$ g cm$^{-3}$, these curves are in good agreement 
with those in Ref.\ \cite{MDH}. 
This observation suggests a possible explanation 
to a discrepancy in occupation numbers of excited states, 
recently recognized \cite{IR95} between OPAL and MDH data: 
the former ones take into account all thermodynamically 
significant while the latter ones only optically identifiable 
atomic states. 

For comparison, long dashes show the solution of the ideal-gas Saha 
equation including the ground state atoms only. These curves 
reproduce accurately the number of ground state atoms at low 
densities. However, at $T\gtrsim 10^{4.5}$~K the total number 
density $n_{\rm H}$ can never be determined in this way, since the 
excited states become populated and increase the neutral fraction. 
Above $10^5$~K, the onset of occupation of the excited states 
produces typical ``shoulders'' on the solid curves. Note that 
highest populated excited states at low densities are strongly 
affected by microfields and belong to the optical quasi-continuum. 
This explains why the low-density tails of the heavy lines lie 
significantly lower than the light ones. 
This difference is noticeable not only 
in the pressure ionization domain $\rho\gtrsim 0.1$ g cm$^{-3}$ 
(where the present model has a limited applicability), 
but also at lower densities, if the temperature 
is high enough for population of the excited states. 
On the other hand, when 
density increases, the pressure ionization comes into effect, the 
neutral fraction becomes smaller and finally disappears at 
$\rho\gtrsim3$ g cm$^{-3}$. There is a considerable amount of bound 
species at $\rho\sim 0.1 - 1$ g cm$^{-3}$ (important for 
thermodynamics) which can hardly contribute to the atomic opacities.
The optical properties of atoms are destroyed
at $\rho > 10^{-2}$ g cm$^{-3}$, as is seen from downward bending
of the heavy lines. 

\begin{figure}[t]
\begin{center}
\epsfysize=60mm 
\epsfbox[70 220 550 540]{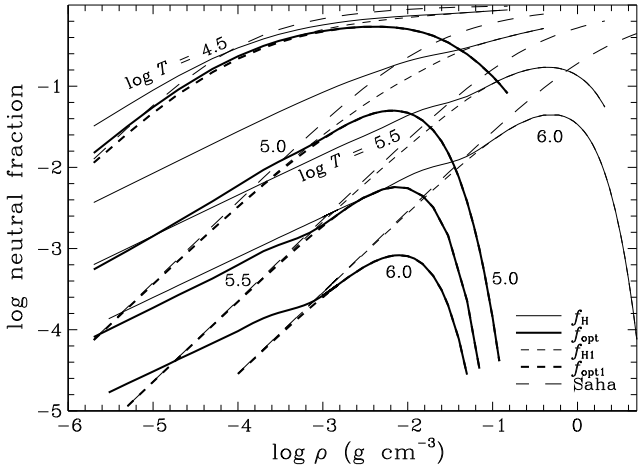}
\end{center}
\caption[]{
Total neutral fraction ($f_{\rm H}$) and partial fractions: 
ground state atoms ($f_{\rm H1}$), optically identifiable bound atoms 
($f_{\rm opt}$), and optically identifiable  
ground state atoms ($f_{\rm opt1}$), compared with the ideal gas 
(Saha) approximation.
}
\label{fig1}
\end{figure}
\subsection{Equation of state}
\label{sect:EOS}
Relative importance of partial contributions to the free energy 
can be estimated by examination of their influence on EOS, 
as illustrated in Fig.~2 for two 
temperatures. At lower temperature, $T=12\,600$~K, the ionization 
degree is small, and the corrections due to the Coulomb nonideality 
of the charged component are practically unimportant. The most 
important corrections are produced by the repulsion of atoms (HS) 
and by the unbounding. The corresponding contributions to the 
pressure are nearly equal and become appreciable at $\rho\gtrsim 
0.03$ g cm$^{-3}$. At higher temperature, $T=2\cdot 10^5$~K, the 
ionization is high, and the Coulomb nonideality is significant, 
while the HS contribution is nearly negligible. The unbounding, 
however, is important at this temperature also. Note that the 
perturbation (van der Waals) terms are unimportant at any $T$ and 
$\rho$ shown in Fig.~2.

\begin{figure}[t]
\begin{center}
\epsfysize=59mm 
\epsfbox[80 220 560 530]{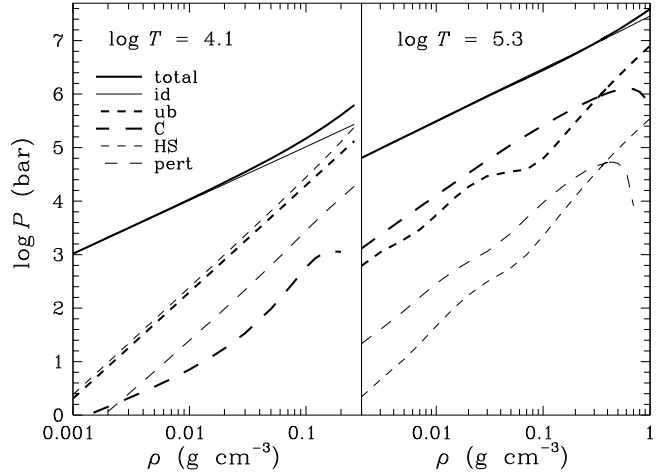}
\end{center}
\caption[]{
Partial pressures: ideal gas (id) and 
configurational parts due to the Coulomb interactions of the ionized 
fraction (C), unbounding of atoms (ub), strong repulsion 
at short distances (HS), and long-distance attraction (pert). 
We show $|P_{\rm C}|$ and $|P_{\rm pert}|$ since these 
parts are negative.
}
\label{fig2}
\end{figure}

In Fig.~3 we compare the present EOS (heavy solid curves) 
with those known in the 
literature. 
Results of FGVH \cite{FGVH} are shown by dashes, 
MDH \cite{MDH} by dot-dashed lines, 
SC \cite{SCVH} by light solid lines, 
and OPAL \cite{OPAL} by dots. The EOS of SC 
is somewhat softer, while that of 
MDH much stiffer (probably owing to the rapidly increasing 
``pressure ionization'' term introduced by the latter authors). We 
have terminated our curves at densities where the quantum correction 
in Eq.~(\ref{4.10}) reached 0.5. At higher $\rho$ the nonideality of 
the charged component becomes too strong to be treated as 
perturbation. The lowest isotherm corresponds to $T=12\,600$~K, 
which is below the critical temperature for the plasma phase 
transition reported by SC \cite{SC92} (the density discontinuity 
seen on the corresponding curve). 

Our results have been obtained under the assumption 
that the molecules are completely destroyed. 
All authors agree that it is true 
at $T > 10^{4.5}$~K, however there is a great 
uncertainty concerning the degree of dissociation 
at $4.1 \lesssim \log_{10} T \lesssim 4.5$. 
According to Ref.\ \cite{MDH}, the amount of molecules 
in this interval is quite insignificant at any density,  
while SC \cite{SC92,SCVH} and 
Reinholz {\it et al.\/} \cite{RRN95} (RRN) 
found similarly strong but quantitatively different 
recombination at density increasing 
from $10^{-2}$ toward 1 g cm$^{-3}$. 
Quantum molecular dynamics simulations \cite{Collins95} 
show that there is a significant amount 
of transient H$_2$-like clusters at $k_{\rm B}T=1$ eV 
and $\rho\sim 1$ g cm$^{-3}$, although their influence on 
thermodynamic and optical properties of plasma 
is not yet well understood. 
\begin{figure}[t]
\begin{center}
\epsfysize=104mm 
\epsfbox[110 177 440 665]{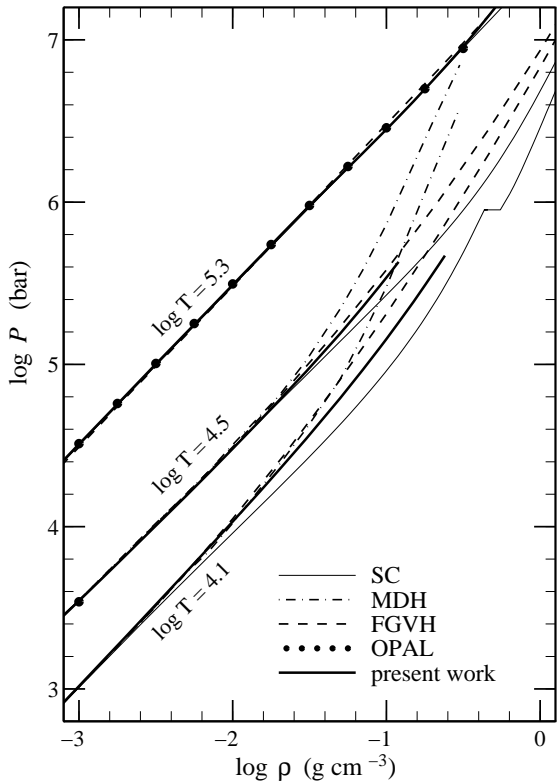}
\end{center}
\caption[]{
Comparison of present EOS with results from the literature.
}
\label{fig3}
\end{figure}
Figure~4 illustrates the effect of this uncertainty 
on the EOS. The pressure isotherm ($T=15\,000$ K) 
corresponding to complete dissociation 
(heavy solid line) is compared with two 
modified isotherms, obtained through replacing 
in our model the 
{\it ideal\/} contribution to $P$ by that 
corresponding to the dissociation 
degree given either by SC (dot-dashed line) 
or by RRN (dashed line). A comparison with the isotherm of SC 
(drawn by light solid line) suggests that the difference 
between our and SC EOS at $\log_{10} T \lesssim 4.5$ 
is mainly caused by the formation of dimers, 
which we neglected. Our testing calculations 
revealed, however, that the dissociation degree 
in the considered $\rho - T$ domain 
depends crucially on treatment 
of various molecular excitations as well as 
on adopted HS diameters. 
Thus the thorny problem of clustering hardly 
can get an unambiguous solution within 
the chemical picture. 

Figure 5 shows the adiabatic temperature gradient 
$\nabla_{\rm ad} = (\partial\ln T/\partial\ln P)_S$, a 
significant quantity sensitive to details of the free energy model. 
Thermal radiation 
starts to dominate in thermodynamics for the hottest isotherm at low 
density, causing the decrease of $\nabla_{\rm ad}$. Other 
depressions of $\nabla_{\rm ad}$ are explained by the increase of 
the specific heat in the  regions of partial ionization, where the 
internal energy is affected by the strongly $T$-dependent ionization 
degree. Our data (heavy curves) are compared with the tables of 
FGVH \cite{FGVH} (dashes), SC \cite{SCVH} 
(dot-dashed lines, ``table'') 
and OPAL \cite{OPAL} (dots). Light solid lines (``formula'') 
are obtained by substituting the SC tabulated quantities 
$P$, $S$, $T$, $a=(\partial\log\rho/\partial\log T)_P$ 
and $b=(\partial\log S/\partial\log T)_P$ 
into the thermodynamic identity 
$\nabla_{\rm ad} = - aPV/(bST)$. 
We have plotted only the latter lines for the two highest 
isotherms, for which the ``table'' and ``formula'' values 
are in a good agreement. 
The discrepancy between these values 
in the low-$T$ high-$\rho$ domain 
indicates the lack of {\it thermodynamic consistency} \cite{SCVH} 
caused by a high sensitivity of the 
complicated numerical approach to accidental small 
errors in minimization and differentiation 
procedures. Although our simplified analytic model 
is less scrupulous in details, its advantage is 
that it is free of such inconsistencies. 

\begin{figure}[t]
\begin{center}
\epsfysize=103mm 
\epsfbox[95 177 300 650]{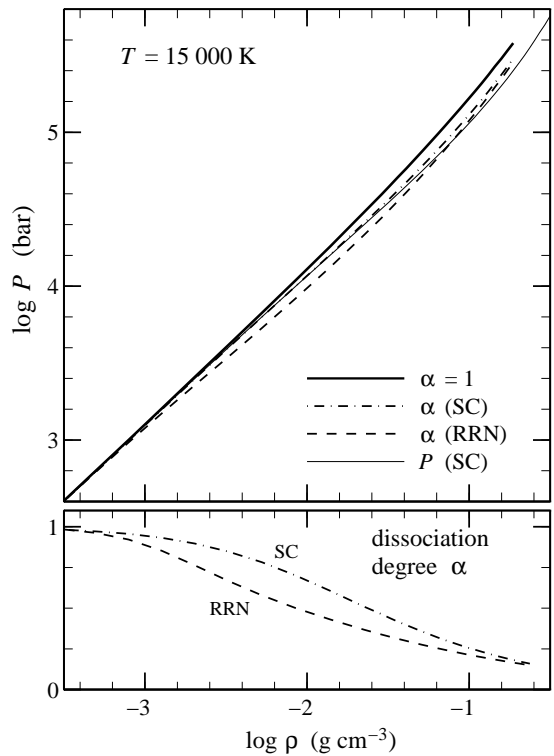}
\end{center}
\caption[]{
Influence of the uncertainty in the dissociation degree 
at high density on the EOS.
}
\label{fig4}
\end{figure}

The most crucial test for the validity of our model may be 
provided by a comparison with the advanced results based on 
the physical picture 
and employed in the OPAL opacity library \cite{OPAL}. 
They do not cover the most interesting 
region where other models reveal major 
discrepancies, but the available data are in satisfactory agreement 
with our model. 

Sequences of shallow depressions in the isotherms $T\ge 
10^5$~K at $\rho>10^{-2}$ g cm$^{-3}$ indicate successive pressure 
destruction of excited atomic states. 
Their physical reality remains an open question. 
Note that the SC results reveal analogous oscillatory 
behavior (which is probably inherent to the models 
of such a type), which however is not observed in the OPAL data. 

\begin{figure}[t]
\begin{center}
\epsfysize=115mm 
\epsfbox[80 145 350 710]{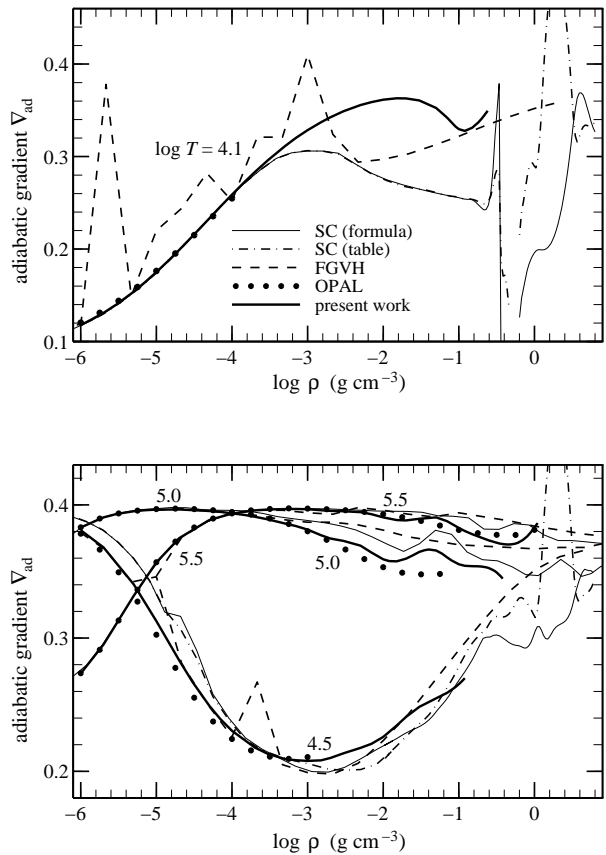}
\end{center}
\caption[]{
Adiabatic temperature gradient compared with the data of FGVH,  
SC, and OPAL. The curves are labeled by $\log_{10}T$ (K).
}
\label{fig5}
\end{figure}
\section{Summary}                                % SUMMARY
\label{sect:sum}
We have developed an analytic free energy model for partially 
ionized hydrogen plasma in the framework of the chemical picture. 
The model describes thermodynamic properties of the plasma at 
$T > 10^4$~K and $\rho \lesssim 0.1 - 1$ g cm$^{-3}$. 
In particular, it can be used in the studies of 
DA white dwarf and neutron star atmospheres. 

The occupation probability formalism, first introduced by Fermi 
\cite{Fermi24} and further developed by Hummer and Mihalas \cite{HM}, is 
generalized to take proper account of the effects of partial 
ionization. Free energy minimization is obtained by a generalized Saha 
equation which is solved by an iterative algorithm involving the 
modified occupation probabilities. Calculated ionization degree 
differs from that obtained previously in Ref.\ \cite{MDH}, but 
qualitatively 
agrees with the results of Ref.\ \cite{SC92}. We argue that the 
relatively high neutral fraction obtained in Ref.\ \cite{SC92} and 
in our present work at $\rho\sim 0.1 - 1$ g cm$^{-3}$ 
cannot be pronounced in atomic opacities, since it takes 
cumulative account of atomic and cluster states, the latter ones 
contributing to the optical quasicontinuum 
(Sect.~\ref{sect:optprob}). We introduce the optical occupation 
probabilities which determine the neutral fraction visible in atomic 
lines. The non-perturbed atomic fraction given by these 
probabilities agrees with that of Ref.\ \cite{MDH} at 
the densities available in laboratory. 

The equation of state obtained from our model is compared with the 
results of other authors. The best agreement is achieved with the 
OPAL data \cite{OPAL} (in the $\rho - T$ region where they are 
available). Since the equation of state employed in OPAL is based on 
the physical picture of plasma, completely different from our model, 
we regard this agreement as an indirect confirmation of the validity 
of our approach. Its generalization to higher densities 
and higher atomic numbers is being performed. 
\begin{acknowledgements}
I am pleased to acknowledge useful discussions with 
G.~G. Pavlov, Y.~A. Shibanov, and D.~G. Yakovlev, 
valuable comments of G.~Chabrier, F.~J. Rogers, and D.~Saumon, 
and consultations with V.~E. Zavlin. 
The hospitality of Gilles Chabrier at the Ecole Normale Sup\'erieure 
de Lyon is gratefully acknowledged. 
The work was supported in part by 
RBRF (Grant 96-02-16870) and INTAS (Grant 94-3834). 
\end{acknowledgements} 
\appendix
\section*{Free energy model based on 
interparticle pair potentials}                           %  Appendix 
Our model correction to the excess free energy $F_{\rm pair}$ 
due to binary interactions among atoms and protons 
is mainly based on Refs.\ \cite{SC91,SC92}. 
Some modifications, however, are introduced to improve the physical 
consistency of the model, and analytic fits are presented. 

The total excess free energy consists of the reference part $F_{\rm 
HS}$, treated in the hard-sphere approximation, and perturbation 
parts $F_{\rm pert}$ required to include the attractive (van der 
Waals) interactions. 
\subsection{Atom-atom interactions} 
\label{sect:neuneu}
We treat interactions among neutral species using the fluid 
perturbation theory of Weeks, Chandler, and Andersen \cite{WCA} 
(WCA). An effective potential $\phi(r)$ is separated 
into the reference and perturbation parts, 
\begin{equation}
    \phi(r) = \phi_{\rm ref}(r) + \phi_{\rm pert}(r), 
\label{4.11}
\end{equation}
where $\phi_{\rm ref}$ is a {\it purely repulsive finite-range\/} 
potential which acts at distances $r<r_\ast$, where $r_\ast$ is 
the minimum point of $\phi(r)$. Accordingly, the free energy 
splits into the reference and perturbation parts. 
First we consider the ground state atoms 
and adopt the interatomic potential 
\begin{equation}
    \phi_{\rm H}(r) = \phi^{({\rm H})}_{\rm SC}(r) + 
    \phi^{({\rm H})}_{\rm core}(r), 
\label{4.16}
\end{equation}
where $\phi^{({\rm H})}_{\rm SC}(r)$ is given by Eq.~(2) of Ref.\ 
\cite{SC91}, and $\phi^{({\rm H})}_{\rm core}(r)$ is a correction 
which acts at short distances $r\lesssim 1.5$~\AA. The correction is 
required to make $\phi(r)$ go to infinity at $r\to 0$. 
Unlike SC \cite{SC92} who introduced HS cores with fixed {\it ad 
hoc\/} diameters, we use the scaled mean electric potential of the 
ground state H atom: 
\begin{equation}
    \phi^{({\rm H})}_{\rm core}(r) = {e^2\over r}\,\left(1+{r\over 
    a}\right){\rm e}^{-2r/a}. 
\label{4.17}
\end{equation}
This choice seems reasonable since it yields the Coulomb 
repulsion at $r \to 0$. We use the effective He screening length 
\cite{BetheSalpeter57} $a= \frac{16}{27} a_{\rm B}$, 
where $a_{\rm B}$ is the Bohr radius, 
for the ground state atom. 

The perturbation part of the free energy is 
\begin{equation}
   F^{({\rm H})}_{\rm pert}/N_{\rm H} = {n_{\rm H}\over 2}
   \int\phi_{\rm pert}(r) g_{\rm ref}(r) {\rm d}^3 r, 
\label{4.19}
\end{equation}
where $g_{\rm ref}(r)$ is the pair correlation function of the 
reference system. For the HS reference system at low densities, 
$g_{\rm ref}(r)$ can be replaced by 0 at $r<d$, and 1 at $r>d$, 
where $d$ is the HS diameter. Then 
\begin{equation}
    F^{({\rm H})}_{\rm pert} / N_{\rm H} = - n_{\rm H}\epsilon_0 
    (2\pi/3)\left(R_0^3-d^3\right), 
\label{4.20}
\end{equation}
where $\epsilon_0/k_{\rm B} = 20.2$~K and $R_0 = 8.6 a_{\rm B}$ for 
the ground state atom.

The diameter $d$ can be determined from the WCA self-consistency 
condition, which involves radial distribution function. However, 
according to Ref.\ \cite{VerletWeis72}, the WCA value of $d$ at low 
$\rho$ is close to the Barker's \cite{BarkerHenderson67} value 
\begin{equation}
    d_{\rm B} = \int_0^\infty\left[1-\exp(-\beta\phi_{\rm ref}(r)) 
    \right]{\rm d} r. 
\label{4.14}
\end{equation}
We adopt $d=d_{\rm B}$ at $\rho\ll1$ g cm$^{-3}$, and propose the 
expression $d=d_{\rm B}\exp(-\rho/2\mbox{~g~cm}^{-3})$ at 
$\rho\lesssim 1$ g cm$^{-3}$. The latter expression fits exact 
numerical results \cite{SC91} with an error of about 5\%. 

We have used the reference part of the potential (\ref{4.16}) in 
Eq.~(\ref{4.14}) and fitted the result by the formula 
\begin{equation}
    d_{\rm H} = d_0 \left[
    1+\ln(1+c_1\sqrt{t})+{t\over 
    1+\ln(1+c_2 t)}\right]^{-1}, 
\label{4.18}
\end{equation}
where $d_0=6a_{\rm B}$, $c_1=4$, $c_2=0.5$, 
and $t=T/(3.25\cdot 10^4$~K) for the ground state atom. 

We use simple scaling of {\it the potentials} for excited states. 
For two atoms with principal quantum numbers $n$ and $n'$, we write 
\begin{equation}
    \phi_{n'n}(r) = s_{n'n}^{-1}\,\phi_{\rm H}(r/s_{n'n}), 
\label{4.20a}
\end{equation}
where $s_{n'n}=(l_n+l_{n'})/(2l_1)$ is the scaling factor, and 
$l_n$ is an average atomic size. The scaling (\ref{4.20a}) ensures 
the correct Coulomb repulsion at short distances. 
We estimate $l_n^2$ as the quantum-mechanical 
expectation value of $r^2$ \cite{LaLi76} 
averaged over the quantum numbers $(l,m)$, 
which yields 
\begin{equation} 
    l_n =  a_{\rm B}\,n\,\sqrt{(7n^2+5)/4} 
\label{4.21} 
\end{equation} 
The scaling does not reduce to just multiplying $d_{\rm H}$ by $s$, 
but implies simultaneous scaling of the temperature: 
\begin{equation}
    d_{nn'}(T) = s_{nn'}\, 
    d_{\rm H}(s_{nn'}T). 
\label{4.21a}
\end{equation}
The perturbation terms for different states are 
additive: 
\begin{equation}
    F_{\rm pert}^{({\rm H})} = - \sum_{\kappa\kappa'} 
    N_\kappa N_{\kappa'} a_{\kappa\kappa'}/V, 
\label{4.21b}
\end{equation}
where the van der Waals constants $a_{\kappa\kappa'}$ are 
determined by the scaled Eq.~(\ref{4.20}), 
\begin{equation}
    a_{\kappa\kappa'} = -s_{\kappa\kappa'}^2\, \epsilon_0 
    \frac{2\pi}{3}
    \left(R_0^3 - d_{\rm H}^3(s_{\kappa\kappa'}T) \right). 
\label{A2}
\end{equation}
\subsection{Ion-atom interactions}         % ION-ATOM INTERACTIONS
\label{sect:ionneu}
Following SC \cite{SC92}, we describe the polarization 
interaction outside the core by the screened dipolar potential: 
\begin{equation}
    \phi_{{\rm pol,}n} = 
    -{e^2\alpha_n\over 2} \left({1+r/ r_{\rm D} \over 
    \l_n^2 + r^2}\right)^2{\rm e}^{- 2 r/ r_{\rm D}}, 
\label{4.22}
\end{equation}
where $\alpha_n$ is the average polarizability of an atom with the 
principal quantum number $n$, and $ r_{\rm D}$ is the screening 
length. The rms size $l_n$ in the denominator of Eq.~(\ref{4.22}) is 
intermediate between two different values of the polarization radius 
used in Ref.\ \cite{SC92}. Furthermore, the 
polarizability of H atom in the state $(nlm)$ averaged over $(lm)$ 
numbers \cite{LaLi76} can be fitted, with an error of 2\%, by 
\begin{equation}
    \alpha_n = 0.85 l_n^3. 
\label{4.23}
\end{equation}
At short separations, the Coulomb repulsion should prevail. 
Therefore, in analogy with Eqs.\ (\ref{4.16}), we adopt 
the interaction potential 
\begin{equation}
    \phi_{\rm in}(r) = \phi_{\rm pol}(r) + 
    \phi^{\rm (in)}_{\rm core}(r), 
\label{4.24}
\end{equation}
where $\phi_{\rm core}$ is given by Eq.~(\ref{4.17}) with 
$a =  \frac12 s_n a_{\rm B}$ and 
the scaling factor $s_n = l_n/l_1$. 

The potential (\ref{4.24}) is separated then into the reference and 
perturbation parts, Eq.~(\ref{4.11}), according to the WCA 
prescription. The perturbation free energy is calculated analogously 
to Eq.~(\ref{4.19}) (with $n_{\rm p}$ instead of $\frac12 n_{\rm H}$ 
on the right-hand side). This gives 
\begin{equation}
     F^{\rm (in)}_{\rm pert} = - \sum_\kappa N_{\rm p} N_\kappa 
      a_\kappa/V, 
\label{4.25}
\end{equation}
where 
\begin{equation}
    a_\kappa = e^2 l_\kappa^2 \left[ v_\ast(l_\kappa/ r_{\rm D}) - 
    {4\pi\over 3} \,\phi_\ast(l_\kappa/ r_{\rm D})\,
    d_\ast^3(s_\kappa T) 
    \right]. 
\label{A4}
\end{equation}
Here $e^2 l_\kappa^2 v_\ast$ is the integral of $\phi^{\rm 
(in)}_{\rm pert}(r)$ over space, $(-e^2/l_\kappa)\phi_\ast$ is the 
minimum of $\phi_{\rm in}$, and $d_\kappa (T) = 
l_\kappa d_\ast (s_\kappa T)$ is the HS diameter determined by 
Eq.~(\ref{4.14}) (nearly independent of $r_{\rm D}$). 
We have obtained the fits 
\begin{equation}
    v_\ast(x) = {4-1.7x\over 1+1.7x}, ~~~
    \phi_\ast(x) = {0.155-0.0212x^2\over 1+0.34x^2};
\label{A6}
\end{equation}
and $d_\ast(T)$ is given by Eq.~(\ref{4.18}) with
$d_0=0.615$, $c_1=0.71$, $c_2=0.75$, and $t=T/(2.15\cdot 10^5$~K). 
\subsection{Hard sphere contribution}          % HS CONTRIBUTION
\label{sect:HS}
The HS diameters depend on atomic states and differ for interactions 
with atoms and ions. Thus we have a {\it non-additive\/} HS mixture.  
Such mixtures can be described by the van der Waals one-fluid model, 
which is reasonably accurate for effective filling factors 
$\eta<0.3$ \cite{Jung-}. In the spirit of this model, we define 
\begin{equation}
    \eta = {\pi\over 6NV}\sum_\kappa N_\kappa \left[ \sum_{\kappa'} 
    N_{\kappa'}d^3_{\kappa\kappa'} + 
    2 N_{\rm p} d^3_{\kappa} \right]
\label{4.27}
\end{equation}
and use the Carnahan--Starling \cite{CarnahanStarling69} 
formula  
\begin{equation}
    \beta F_{\rm HS}/N = (4\eta-3\eta^2)/(1-\eta)^2. 
\label{4.28}
\end{equation}
Finally, the excess free energy associated with 
the pair potentials is given 
by Eqs.\ (\ref{4.28}), (\ref{4.21b}) and (\ref{4.25}): 
\begin{equation}
    F_{\rm pair} = F_{\rm HS} + F^{({\rm H})}_{\rm pert} + 
    F^{\rm (in)}_{\rm pert}. 
\label{A1}
\end{equation}


\begin{references}
\bibitem{Ediss}
In Eq.~(20) of Ref.\ \cite{RRN95}, 
the spectroscopic constant 4.75 eV should be replaced 
by the dissociation energy 4.48 eV, and 
the factors $\beta h c$ should be inverted 
(e.g., Ref.~\cite{LaLi86})
\end{references}
\end{document}